\documentclass[runningheads,a4paper]{llncs}

\usepackage[T1]{fontenc}
\usepackage[utf8]{inputenc}
\usepackage{amssymb}
\usepackage{graphicx}
\usepackage{rotating}
\usepackage{array}
\usepackage{amsmath}
\usepackage{url}

\usepackage{comment}
\usepackage[linesnumbered,ruled,vlined]{algorithm2e}
\urldef{\mailsc}\path|contact@alexandrebazin.com,{jcarbonnel,huchard}@lirmm.fr, giacomo.kahn@isima.fr|

\newcommand{\keywords}[1]{\par\addvspace\baselineskip
\noindent\keywordname\enspace\ignorespaces#1}

\begin{document}

\mainmatter  % start of an individual contribution

\title{On-demand Relational Concept Analysis}

\titlerunning{On-demand Relational Concept Analysis}

\author{Alexandre Bazin \inst{1}
\and Jessie Carbonnel \inst{2}
\and Marianne Huchard \inst{2}
\and Giacomo Kahn \inst{3}
 }
\authorrunning{Bazin, Carbonnel, Huchard, Kahn}

\institute{
Le2i, Université Bourgogne Franche-Comté, Dijon, France\\
\and LIRMM, CNRS and Université de Montpellier, Montpellier France\\ %\\
%161, rue Ada, 34095 Montpellier cedex 5, France \\
%\url{http://www.lirmm.fr}}
\and LIMOS, Université Clermont Auvergne, Clermont-Ferrand, France\\
\mailsc}
%Av. Michel Crépeau
%17042 La Rochelle Cedex 1, France

\toctitle{Lecture Notes in Computer Science}
\tocauthor{Authors' Instructions}
\maketitle

\begin{abstract}
Formal Concept Analysis and its associated conceptual structures have been used to support exploratory search through conceptual navigation.
Relational Concept Analysis (RCA) is an extension of Formal Concept Analysis to process relational datasets.
RCA and its multiple interconnected structures represent good candidates to support exploratory search in relational datasets, as they are enabling navigation within a structure as well as between the connected structures.
However, building the entire structures does not present an efficient solution to explore a small localised area of the dataset, for instance to retrieve the closest alternatives to a given query.
In these cases, generating only a concept and its neighbour concepts at each navigation step
 appears as a less costly alternative.
In this paper, we propose an algorithm to compute a concept and its neighbourhood in extended concept lattices.
The concepts are generated directly from the relational context family, and  possess both formal and relational attributes.
The algorithm takes into account two RCA scaling operators.
We illustrate it on an example.

\end{abstract}

\section{Introduction}
\label{sec:intro}

Many datasets in thematic areas like environment or product lines comprise databases complying with a relational data model. Typical applications in which we are currently involved concern issues relative to  watercourse quality\footnote{http://engees-fresqueau.unistra.fr/presentation.php?lang=en} (Fresqu\-eau project), the inventory and use of pesticidal, antibacterial and antifungal  plants\footnote{http://www.cirad.fr/en/news/all-news-items/articles/2017/science/identifying-plants-used-as-natural-pesticides-in-africa-knomana} (Knomana project),
and the analysis and representation of product lines \cite{Carbonnel2017Analyzing}.
In these applications, there is a wide range of question forms, such as classical querying, establishing correlations between descriptions of objects from several categories or case based reasoning.
These questions can be addressed by complementary approaches including conceptual classification building, knowledge pattern and rule extraction, or exploratory search \cite{DBLP:journals/cacm/Marchionini06,DBLP:conf/iui/PalagiGGT17}.
In the Knomana project, for example, one main purpose will be, after the ongoing inventory, to support farmers, their advisors, local entrepreneurs or researchers in selecting plants of immediate interest for agricultural crop protection and animal health. As such users will face large amounts of data, and mainly will formulate general, potentially imprecise, and potentially inaccurate queries without prior knowledge of the data, exploratory search will be a suitable approach in this context.

Previous work \cite{Godin1986Browsable,Carpineto2004,Ducrou2007,Ferre14Reconciling,DBLP:journals/scientometrics/DunaiskiG017} has shown that Formal Concept Analysis may be a relevant support for data exploration and we expect Relational Concept Analysis (RCA) to be beneficial as well. Considering RCA for relational dataset exploration brings issues relative to the use of the scaling (logical) operators, the iterative process and the presence of several concept lattices connected via relational attributes. Despite this additional complexity, RCA helps the user to concentrate on the classification of objects of several categories, where the object groups (concepts) are described by intrinsic attributes and by their relations to object groups of other categories. Besides, the relational attributes offer a support to navigate between the object groups of the different categories, while the concept lattices offer a (by-specialisation) navigation between object groups of the same category.

There are several complementary
%tracks
strategies to explore datasets using RCA. One may consist in exhaustively computing concept lattices (and related artefacts like implication rules) at several steps, using several logical operators and considering only some of the object categories and some of the inter-categories relationships. Another strategy, which is followed here, consists in an on-demand computation of a concept and its neighbourhood comprising its upper, lower and relational covers.

The next section presents the main principles of Relational Concept Analysis (Section \ref{sec:rca}).
The on-demand computation of a concept and its neighbourhood is presented in Section \ref{sec:algorithms}.
 Section \ref{sec:examples} illustrates the algorithm with the example introduced in Section \ref{sec:rca}.
 Related work is exposed in Section \ref{sec:relatedwork}. We conclude the paper with a few perspectives in Section \ref{sec:conclusion}.
 %2 pages with abstract and title
%\input{motivations}%2 pages
\section{Relational Concept Analysis}
\label{sec:rca}

%%%%% FCA basics
Formal Concept Analysis (FCA) \cite{Ganter1999Formal} allows to structure a set of objects described by attributes in a canonical structure called a concept lattice.
It is based on a formal context $K = (\mathcal O,\mathcal A,\mathcal I)$, where $\mathcal O$ is the set of objects, $\mathcal A$ the set of attributes, and $\mathcal I$ an incidence relation stating "which objects possess which attributes".
From this context, the application of FCA extracts a finite set $C_K$ of formal concepts $(X,Y)$
such that $X = \{o \in \mathcal O~|~\forall a \in Y, (o,a) \in \mathcal I\}$
 is the concept's extent, and
 $Y = \{a \in \mathcal A ~|~ \forall o \in X, (o,a) \in\mathcal I\}$
 is the concept's intent.
The concept lattice is obtained by ordering the concepts of $C_K$ by the set-inclusion order on their extents.
We call an \textit{object-concept} (resp. \textit{attribute-concept}) the lowest (resp. the greatest) concept in the lattice possessing an object (resp. an attribute).
%%%%% RCA introduction

Relational Concept Analysis (RCA) \cite{Rouane2007Proposal,Huchard2007Relational} is an adaptation of FCA to process relational datasets.
A relational dataset is composed of several sorts of objects described by both their own attributes and their relationships with other objects.
%%%%% RCF
As input, RCA takes a Relational Context Family (RCF), gathering a set of formal contexts
%representing the different sort of objects and their attributes
 and a set of relational contexts defining links between the objects of different formal contexts.

\begin{definition}[Relational Context Family]
A Relational Context Family is a pair $(\textbf{K},\textbf{R})$ such that:\\
 - $\textbf{K} = \{K_i = (\mathcal O_i,\mathcal A_i,\mathcal I_i)\}$ is a set of formal contexts (object-attribute relations)\\
   - $\textbf{R} = \{r_k\}, r_k \subseteq \mathcal O_i \times \mathcal O_j$ is a set of relational contexts (object-object relations), with $\mathcal O_i$ and $\mathcal O_j$ being sets of objects (respectively of $K_i$ and $K_j$). $K_i$ is called the source context and $K_j$ the target context.
   %$(o_i,o_j) \in r_k$ represents a link between the objects $o_i$ and $o_j$.
\end{definition}

%%%%% Example of RCF
The three contexts of Table~\ref{table:rcf-formalcontexts} present an example of RCF $(\textbf{K}_s,\textbf{R}_s)$ taken from the software product line domain.
Table~\ref{table:rcf-formalcontexts} (top) displays two formal contexts.
The one on the left-hand side presents 5 Data Modelling tools (\textit{DM\_tools}) against 7 attributes representing their compatible operating systems (\textit{OS:}), and the data models (\textit{DM:}) the tools may manage.
The table on the right-hand side describes 4 DataBase Management Systems (\textit{DBMS}) according to the data types (\textit{DT:}) they may handle.
Table~\ref{table:rcf-formalcontexts} (bottom) presents a relational context
%\textit{DM\_tools} $\times$ \textit{DBMS}
 stating which Data Modelling tools \textit{support} which DataBase Management Systems.

\begin{table}[ht]
  \scriptsize
  \caption{(top) Two formal contexts: (left-hand side) Data Modelling tools (DM\_tools) and (right-hand side) DataBase Management Systems (DBMS). (bottom) Relational context stating which DM\_tools \textit{support} which DBMS}
  \label{table:rcf-formalcontexts}
  \centering
{\small  $\textbf{K}_s=$}
  \begin{tabular}{|l|c|c|c|c|c|c|c|}
    \hline
    \textit{DM\_tools}&\begin{sideways}OS:Windows\end{sideways}&\begin{sideways}OS:Mac OS\end{sideways}&\begin{sideways}OS:Linux\end{sideways}&\begin{sideways}DM:Conceptual~~\end{sideways}&\begin{sideways}DM:Physical\end{sideways}&\begin{sideways}DM:Logical\end{sideways}&\begin{sideways}DM:ETL\end{sideways}\\
    \hline
    Astah&x&x&x&x&&&\\
    \hline
    Erwin DM&x&&&x&x&x&\\
    \hline
    ER/Studio&x&&&x&x&x&x\\
    \hline
    Magic Draw&x&x&x&x&x&x&\\
    \hline
    MySQL Workbench&x&x&x&&x&&\\
    \hline
    \end{tabular}
    \begin{tabular}{|l|c|c|c|c|c|c|c|c|c|c|}
      \hline
      \textit{DBMS}&\begin{sideways}DT:Enum~~\end{sideways}&\begin{sideways}DT:Set~\end{sideways}&\begin{sideways}DT:Geometry~~\end{sideways}&\begin{sideways}DT:Spatial\end{sideways}&\begin{sideways}DT:Audio\end{sideways}&\begin{sideways}DT:Image\end{sideways}&\begin{sideways}DT:Video\end{sideways}&\begin{sideways}DT:XML\end{sideways}&\begin{sideways}DT:JSON\end{sideways}&\begin{sideways}DT:Period\end{sideways}\\
      \hline
      MySQL&x&x&x&&&&&&&\\
      \hline
      Oracle&&&&x&x&x&x&x&&\\
      \hline
      PostgreSQL&x&&x&&&&&x&x&\\
      \hline
      Teradata&x&&x&&&&&x&x&x\\
      \hline
      \end{tabular}\\
      \smallskip
      {\small  $\textbf{R}_s=$}
      \begin{tabular}{|l|c|c|c|c|}
        \hline
        \textit{support}&MySQL&Oracle&PostgreSQL&Teradata\\
        \hline
        Astah&x&x&&\\
        \hline
        Erwin DM & x&x&&x\\
        \hline
        ER/Studio&x&x&x&x\\
        \hline
        Magic Draw&x&x&x&\\
        \hline
        MySQL Workbench&x&&&\\
        \hline
        \end{tabular}
  \end{table}

%  \begin{table}[ht]
%    \scriptsize
%    \caption{Relational context stating which DM\_tools \textit{support} which DBMS}
%    \label{table:rcf-relationalcontext}
%    \centering
%    \begin{tabular}{|l|c|c|c|c|}
%      \hline
%      \textit{support}&MySQL&Oracle&PostgreSQL&Teradata\\
%      \hline
%      Astah&x&x&&\\
%      \hline
%      Erwin DM & x&x&&x\\
%      \hline
%      ER/Studio&x&x&x&x\\
%      \hline
%      Magic Draw&x&x&x&\\
%      \hline
%      MySQL Workbench&x&&&\\
%      \hline
%      \end{tabular}
%    \end{table}

\begin{figure}
  \centering
  \includegraphics[width=.8\linewidth]{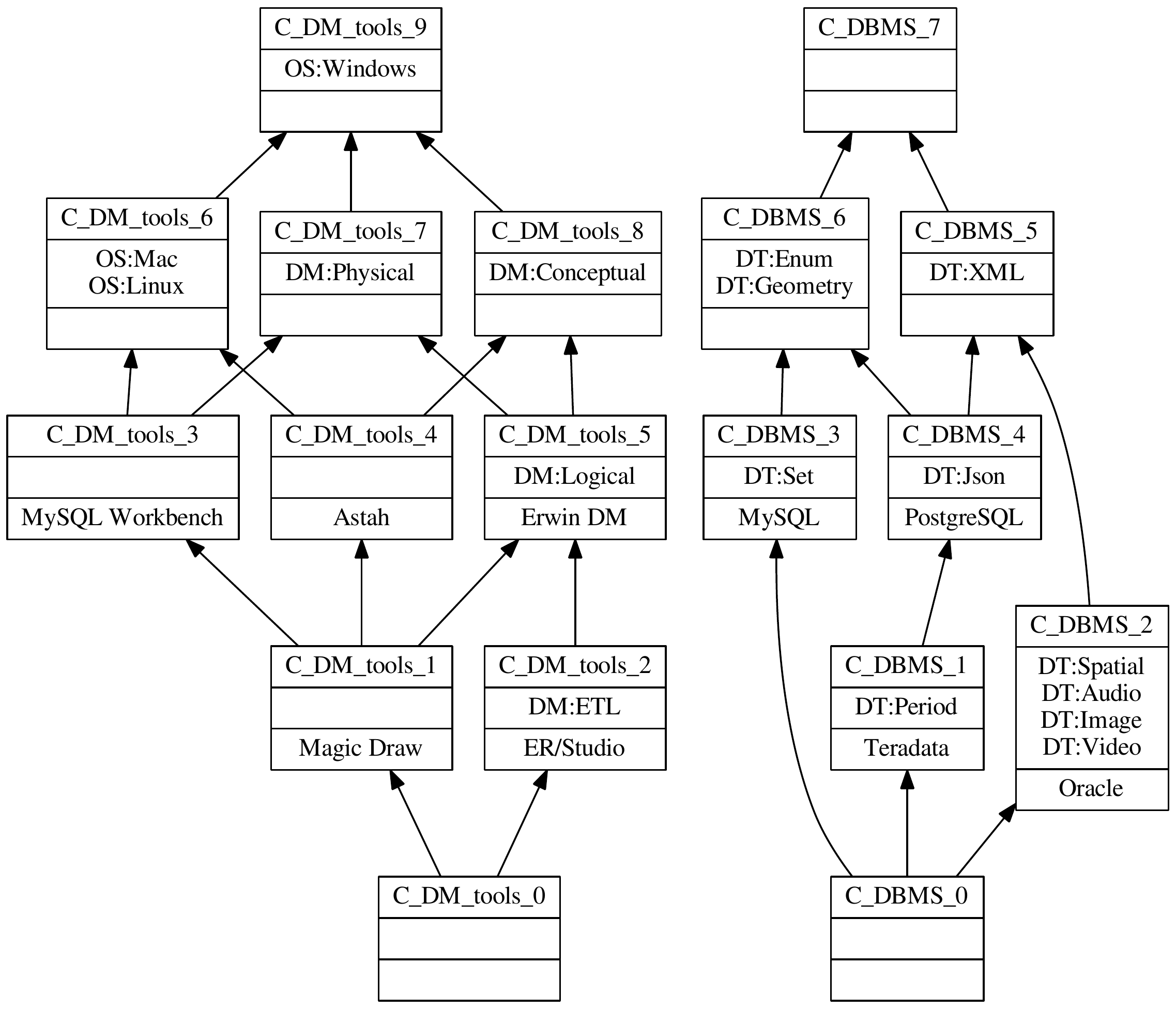}
  \caption{(left) concept lattice of DM\_tools, (right) concept lattice of DBMS}
  \label{fig:conceptlattices}
\end{figure}

Applying RCA on the contexts of $\textbf{K}$ builds, in a first time, one concept lattice per context (i.e., sort/category of objects), without taking links into account.
% where objects are organised depending on their attributes as for traditional FCA.
The two concept lattices associated with Table~\ref{table:rcf-formalcontexts} (top) are presented in Fig.~\ref{fig:conceptlattices}.
In a second time, RCA introduces links between objects of different lattices depending on the relations expressed in $\textbf{R}$.
%Therefore, an object presents in a concept of $K_i$ may be connected with an object presents in another concept from $K_j$.
%%%%% Relational attributes
These links take the form of \textit{relational attributes};
they introduce the abstractions (i.e., concepts) from the target context into the source context through a specific relation and a specific \textit{scaling operator}.
In our example, we may introduce the relational attribute $\exists~support.(C\_DBMS\_4)$ to characterise the \textit{DM\_tools} that support at least one \textit{DBMS} offering Json and XML.
%In our example, they will reference the 8 concepts of \textit{DBMS} in the context \textit{DM\_tools}.
%Given a formal context $F = (O,A,I)$, we define $rel(F) \subseteq \textbf{R}$ the set of relational context having $O$ as source context.
More generally, given two formal contexts $K_i, K_j \in \textbf{K}$ and a relational context $r \subseteq \mathcal O_i \times \mathcal O_j$, the application of RCA extends the set of attributes $\mathcal A_i$ with a set of relational attributes representing links to the concepts of $K_j$.
The extended attribute set is denoted $\mathcal A_i^+$.
%Each added relational attribute represents one concept of the target context $K_j = (O_j,A_j,I_j)$.
Then, the incidence relation $\mathcal I_i$ is extended to take into account these new attributes (denoted $\mathcal I_i^+$), by associating them to each object of $\mathcal O_i$ depending on the relation $r$, the concept (denoted $C$) involved in the relational attribute and a scaling operator $\rho$.
%%%%% Scaling Operator
%This association depends on a \textit{scaling operator} denoted $\rho$, and the concept's extent represented by the relational attribute.
A relational attribute is thus of the form $"\rho~r.(C)"$. %, where $\rho$ is the chosen scaling operator, $r_k \subseteq \mathcal O_i \times \mathcal O_j$ is a relational context, and $C$ is a concept of $K_j$.
In this paper, we focus on two scaling operators:
%\begin{itemize}
  the \textit{existential} operator (denoted $\exists$), associating an object $o$ to the relational attribute $\exists r.(C)$ if $o$ is linked to at least one object of the extent of $C$ by $r$;
%  \item \textit{Contains strict} (denoted ${\exists}{\supseteq}$): an object $o$ is associated to the relational attribute ${\exists}{\supseteq} r_k(C)$ if all the objects in the extent of $C$ are linked to $o$ by $r_k$.
%  \item \textit{Contains percent} (denoted ${\exists}{\supseteq}\ge_{X\%}$): an object $o$ is associated to ${\exists}{\supseteq}\ge_{X\%} r_k(C)$ if at least $X$ percent of the objects in the extent of $C$ are linked to $o$ by $r_k$.
  the \textit{universal strict} operator (denoted $\exists\forall$), associating an object $o$ to $\exists\forall r.(C)$ if all the objects linked to $o$ by $r$ are included in the extent of $C$, and $r(o)\not = \emptyset$.

\begin{figure}
  \centering
  \includegraphics[width=.5\linewidth]{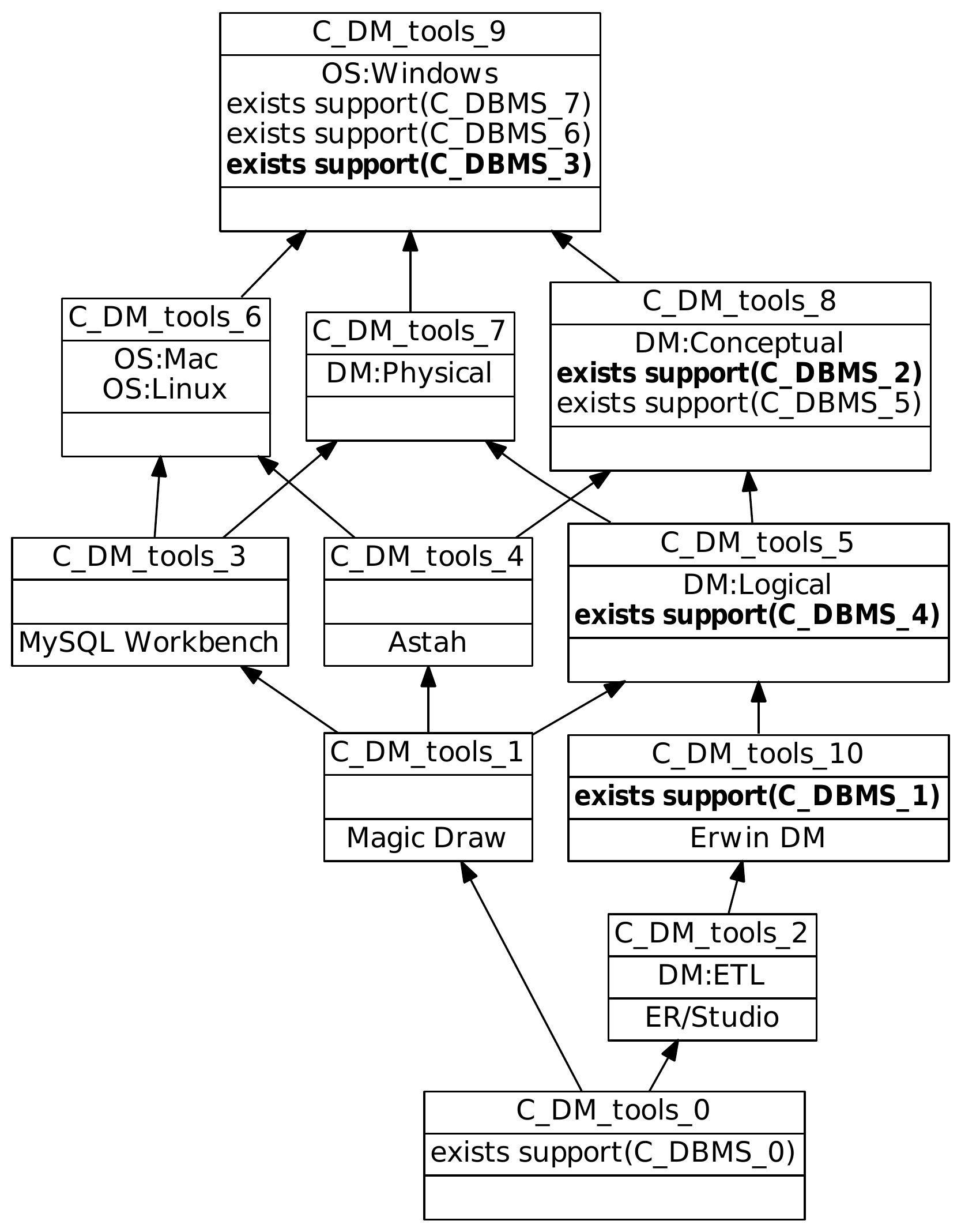}
  \caption{Concept lattice of the extended context \textit{DM\_tools}$^+$ }%of Table~\ref{table:rcf-extendedformalcontext}}
  \label{fig:extendedconceptlattice}
\end{figure}

The concept lattice associated with a formal context $K^+ = (\mathcal O,\mathcal A^+,\mathcal I^+)$ then structures the objects from $\mathcal O$ both by their attributes and their relations to other sets of objects through the relational attributes.
Fig.~\ref{fig:extendedconceptlattice} presents the extended concept lattice corresponding to the extended formal context \textit{DM\_tools}$^+$, according to the relation \textit{support} and the \textit{existential} scaling operator.

In this way, for complex data models including more than one relation, RCA produces a succession of concept lattices, extended at each step by the new abstractions obtained at the previous step.
At step 0, the concept lattices in the set $\textbf{L}_0$ are the ones built from the initial formal contexts from $\textbf{K}$.
At step $n$, the formal contexts in the set $\textbf{K}_n$ are extended depending on the concepts of the concept lattices in $\textbf{L}_{n-1}$ and the relations expressed in $\textbf{R}$.
%The RCF may contain loops, but it always reaches a fix-point \cite{Hacene2013Soundness}.
%2pages
\section{The Exploration Algorithm}
\label{sec:algorithms}

In this section, we present an algorithm for taking a step in an exploration. It considers an RCF potentially extended at previous steps, a starting concept $C$ from a context $\mathcal K_i$ of the RCF and an exploration \textit{strategy} which consists in choosing a set of relations of the RCF (with $\mathcal K_i$ as a source) provided with scaling operators.
The objective of one step is to complete the intent corresponding to the extent of $C$, as well as compute its upper, lower and relational covers. Meanwhile, the RCF is updated with the relational attributes for a next step.

\paragraph{Redefining Derivation Operators}

The explicit knowledge of all the
 relational attributes of a context requires the computation of all the concepts of the target contexts.
However, we cannot afford what amounts to the exhaustive computation of the relational concepts of multiple contexts.
We would prefer to manipulate only a minimal number of relational attributes allowing us to derive, on-the-fly, the other relational attributes.

Any object described by an attribute $\rho ~r.(X,Y)$ (instead of $\rho~r.((X,Y))$ by abuse of notation) is also necessarily described by all the attributes of the form $\rho~ r.(X_2,Y_2)$ where $Y_2\subseteq Y$.
%The same holds for the $\exists\forall$ scaling operator.
As such, intents can be represented without loss of information by their relational attributes constructed from attributes-wise maximal concepts.
However, a problem arises with such a representation: the set intersection cannot be used to compute the intent of a set of objects anymore.
Similarly, if only maximal relational attributes are explicitly present in the context, the extent of a set of attributes cannot be computed through a simple test of set inclusion.
To remedy this, we provide three algorithms to use on sets of attributes (both intrinsic and relational) with only the maximal relational attributes given explicitly.
%\begin{itemize}
%\item {\sc Intersect}, that computes the intersection of two sets of attributes
%\item {\sc In}, that computes the intent of a set of objects
%\item {\sc Ex}, that computes the extent of a set of attributes
%\end{itemize}

{\sc Intersect} takes as input two sets of attributes $A$ and $B$ represented by their maximal relational attributes. It outputs the set of maximal relational attributes of their intersection. A relational attribute $\exists r.(X,Y)$ is in the intersection of $A$ and $B$ if and only if there exists two attributes $\exists r.(X_2,Y_2)\in A$ and $\exists r.(X_3,Y_3)\in B$ such that $X\subseteq X_2$ and $X\subseteq X_3$. The same holds for the $\exists\forall$ scaling operator. As such, intersecting the intents of the concepts in the attributes of $A$ and $B$ and keeping the maximal ones results in the maximal relational attributes. It uses {\sc Ex} (Algorithm \ref{algo:Extent}).

%A garder
\begin{algorithm}[ht]
\DontPrintSemicolon
\KwIn{$\mathcal K_i = (\mathcal O_i, \mathcal A_i, \mathcal I_i)$ a formal context, $A \subseteq \mathcal A_i$ an attribute set, $B \subseteq \mathcal A_i$ the intent of an object $o$}
\KwOut{The relational intersection of the attribute set $A$ and the intent of $o$}
	$A_2\gets A\cap B$\;
	$\mathcal F\gets\emptyset$\;
	\ForEach{$a_1\sim\exists r.(X_1,Y_1)\in B$ with $r\subseteq \mathcal O_i \times \mathcal O_j$ and $\mathcal K_j = (\mathcal O_j, \mathcal A_j, \mathcal I_j)$}{
		\ForEach{$a_2\sim\exists r.(X_2,Y_2)\in A$}{
			$\mathcal F\gets\mathcal F\cup\{\exists r.(${\sc Ex}$(\mathcal K_j,${\sc Intersect}$(\mathcal K_j, Y_1, Y_2)),${\sc Intersect}$(\mathcal K_j,Y_1, Y_2))\}$\;
		}
	}
	$A_2\gets A_2\cup${\sc Max}$(\mathcal F,\subseteq_{\mathcal A_i})$\;
      $\mathcal F\gets\emptyset$\;
	\ForEach{$a_1\sim\exists\forall r.(X_1,Y_1)\in B$ with $r\subseteq \mathcal O_i \times \mathcal O_j$ and $\mathcal K_j = (\mathcal O_j, \mathcal A_j, \mathcal I_j)$}{
		\ForEach{$a_2\sim\exists\forall r.(X_2,Y_2)\in A$}{
			$\mathcal F\gets\mathcal F\cup \{\exists\forall r.(${\sc Ex}$(\mathcal K_j$,{\sc Intersect}$(\mathcal K_j,Y_1, Y_2)),${\sc Intersect}$(\mathcal K_j,Y_1, Y_2))\}$\;
		}
	}
    $A_2\gets A_2\cup\mathcal F$\;
\Return{$A_2$}
\caption{{\sc Intersect}$(\mathcal K_i, A,B)$}
\label{algo:Intersect}
\end{algorithm}

{\sc In} uses {\sc Intersect} to compute the intent of a set of objects described by their maximal relational attributes.
It starts with the set of all explicitly known attributes and intersects it with the description of each object in the context $\mathcal K_i$.

%Pas forcément à garder
\begin{algorithm}[ht]
\DontPrintSemicolon
\KwIn{$\mathcal K_i = (\mathcal O_i, \mathcal A_i, \mathcal I_i)$ a formal context, $O \subseteq \mathcal O_i$ a set of objects}
\KwOut{Computes the intent of a set of objects $O$}

$A\gets \mathcal{A}_{i}$\;
\ForEach{$o\in O$}{
$A\gets ${\sc Intersect}$(A,Intent(\{o\}))$\;
}
\Return A\;

\caption{{\sc In}$(\mathcal K_i, O)$}
\label{algo:Intent}
\end{algorithm}

{\sc Ex} computes the extent of a set of maximal relational attributes A. For each object $o$ and attribute $\rho~ r.(X,Y)\in A$, it checks whether $r(o)$ and $X$ intersect in the correct way (depending on the scaling operator).

%A garder
\begin{algorithm}[ht]
\DontPrintSemicolon
\KwIn{$\mathcal K_i = (\mathcal O_i, \mathcal A_i, \mathcal I_i)$ a formal context, $A \subseteq \mathcal A_i$ a set of attributes}
\KwOut{Computes the extent of a set of attributes $A$}

$O\gets \mathcal{O}_{i}$\;
\ForEach{$a\in A$}{
	\If{$a\sim \exists\forall r.(X,Y)$}{
		\ForEach{$o\in O$}{
        	\If{$r(o)\not\subseteq X$}{
        		$O\gets O\setminus o$\;
			}
        }
    }
    \ElseIf{$a\sim \exists r.(X,Y)$}{
    	\ForEach{$o\in O$}{
        	\If{$r(o)\cap X=\emptyset$}{
        		$O\gets O\setminus o$\;
			}
        }
    }
	\Else{
		\ForEach{$o\in O$}{
        	\If{$(o,a)\not\in\mathcal I_i$}{
        		$O\gets O\setminus o$\;
			}
        }
	}
}
\Return O\;

\caption{{\sc Ex}$(\mathcal K_i, A)$}
\label{algo:Extent}
\end{algorithm}

\paragraph{Computing the Closed Neighbourhood}

Now that we have redefined the derivation operators on implicitly known relational contexts, we are able to compute the upper, lower and relational covers of a concept.

The easiest are the relational covers. A concept $(X,Y)$ is a relational cover of a concept $(U,V)$ if and only if $\rho~ r.(X,Y)$ is a maximal relational attribute in $V$. Upper covers are easy too. Candidates can be generated by adding an object -- the set of which we have perfect knowledge of -- to the current extent and computing the corresponding concept. The covers are the candidates that have the smallest extent. Computing the lower covers is more challenging. They could be computed by adding attributes to the intent but the full set of relational attributes is only known implicitly. We chose to, instead, remove objects. The lower covers of $(X,Y)$ being the concepts with the maximal extents that are contained in $X$ and do not contain any of the minimal generators of $X$, a simple way to compute them would be to remove minimal transversals of the minimal generators.

Algorithm~\ref{algo:RCA} computes the closed neighbourhood of a concept $C$.
It takes as input a set of formal contexts $\textbf{K}=(\mathcal K_1,\dots,\mathcal K_w)$ of a RCF, a strategy $\mathcal S=\{(r,\rho)_{lj},\dots\},$ $ l,j\in \{1,\dots,w\}$ and a starting concept $C$ from a context $\mathcal K_i$.
The goal is to compute (or complete) the intent corresponding to the extent of $C$, as well as its upper, lower and relational covers, in the extended context $\mathcal K^+_i$.

For each $(r,\rho)_{ij}\in\mathcal S$ such that $r : \mathcal K_i\mapsto \mathcal K_j$, the first loop (Lines 1 to 4): computes $OC_j$ the object-concepts of $\mathcal K_j$; then, each object-concept $(X,Y) \in OC_j$, relation $r$ and scaling operator $\rho$ give rise to a new relational attribute $\rho~ r.(X,Y)$ that is added to the context $\mathcal K_i$ with {\sc GrowContext}.

In Line~\ref{update:A}, the intent of concept $C$ is extended with the relational attributes added during the previous loop.
The next loop (Lines~\ref{R:d} to~8) computes the relational covers $\mathcal R$ of concept $C$. For each relational attribute in the intent of $C$, the corresponding concept (in the target context) is added to the cover.

In Lines~\ref{L:d} to~11, the lower covers $\mathcal L$  of $C$ are computed by removing from the extent of $C$ a minimal transversal of the set of minimal generators of $C$'s extent.

Finally, the upper covers $\mathcal U$  of $C$ are computed in Lines~\ref{U:d} to~14. Candidates are created by adding an object $o$ to the extent of $C$. Only the extent-wise minimal resulting concepts are kept.

%A garder
\begin{algorithm}[ht]
\DontPrintSemicolon
\KwIn{$\textbf{K}=\{\mathcal K_1,\dots,\mathcal K_w\}$, $\mathcal S=\{(r,\rho)_{lj},\dots\}, l,j\in \{1,\dots,w\}$ a strategy, $C=(O,A)$ a concept of $\mathcal K_i = (\mathcal O_i, \mathcal A_i, \mathcal I_i)$}
\KwOut{$C, \mathcal U, \mathcal R, \mathcal L$ the completed concept $C$ and its closed relational neighbourhood}

    \ForEach{$(r,\rho)_{ij}\in\mathcal S$}{\label{loop:relational}
    	$OC_j\gets${\sc ObjectPoset}$(\mathcal K_j)$\;

        \ForEach{$o\in\mathcal O_i$}{
			{\sc GrowContext}$(\mathcal K_i,r,\rho,o,OC_j)$\;
		}

    }

	$A\gets${\sc In}$(\mathcal K_i,O)$\;\label{update:A}
    $\mathcal R\gets\emptyset$\;\label{R:d}%debut relational cover
    \ForEach{$a\sim \rho r.(X_1,Y_1)\in A$}{
    	$\mathcal R\gets\mathcal R\cup \{(X_1,Y_1)\}$\;
    }\label{R:f}%fin relational cover

	$\mathcal L\gets\emptyset$\;\label{L:d}%debut lower cover
    \ForEach{$T\in minTrans(minGen(O))$}{\label{loop:lower}
			$\mathcal L\gets\mathcal L\cup \{(O\setminus T,${\sc In}$(\mathcal K_i,O\setminus T))\}$\;
	}\label{L:f}

	$\mathcal U\gets\emptyset$\;\label{U:d}%debut upper cover
	\ForEach{$o\in\mathcal O_i\setminus O$}{
		$\mathcal U\gets\mathcal U\cup\{(${\sc Ex}$(\mathcal K_i,${\sc In}$(\mathcal K_i,O\cup \{o\}))$, {\sc In}$(\mathcal K_i,O\cup \{o\}))\}$\;
	}
	$\mathcal U\gets ${\sc Min}$(\mathcal U,\subseteq_{\mathcal O_i})$\;\label{U:f}%fin upper cover

\Return{$C, \mathcal U, \mathcal R, \mathcal L$}\;
\caption{{\sc RCA}$(\textbf{K}, \mathcal S, C,\mathcal K_i)$}
\label{algo:RCA}
\end{algorithm}

%Pas forcément à garder
%\begin{algorithm}[ht]
%\DontPrintSemicolon
%\KwIn{$\mathcal K_i$ a formal context, $r \subseteq \mathcal O_i \times \mathcal O_j$ a relational context, $\rho$ a scaling operator, $o \in \mathcal O_i$ an object, $OC_j$ a set of object-concepts of $\mathcal K_j$}
%\KwOut{Extends the context $\mathcal K_i$ and adds the crosses}
%\If{$\rho==\exists$}{
%	\ForEach{$obj\in r(o)$}{
	%	$\mathcal A_i\gets\mathcal A_i\cup \exists r.(OC_j(obj))$\;
		%$\mathcal I_i\gets\mathcal I_i\cup (o, \exists r.(OC_j(obj)))$\;
	%}
%}

%\If{$\rho==\exists\forall$}{
	%$X\gets\mathcal A_j$\;
	%\ForEach{$obj\in r(o)$}{
	%	$X\gets ${\sc Intersect}$(X, ${\sc ~In}$(\mathcal K_j,OC_j(obj)))$\;
%    }
%    $\mathcal A_i\gets\mathcal A_i\cup \exists\forall r.(${\sc Ex}$(\mathcal K_i, X), X)$\;
%	$\mathcal I_i\gets\mathcal I_i\cup (o, \exists\forall r.(${\sc Ex}$(\mathcal K_i,X), X))$\;
%}
%\caption{{\sc GrowContext}$(\mathcal K_i,r,q,o,OC_j)$}
%\label{algo:Extendsold}
%\end{algorithm}

\begin{algorithm}[ht]
\DontPrintSemicolon
\KwIn{$\mathcal K_i = (\mathcal O_i, \mathcal A_i, \mathcal I_i)$ a formal context, $r \subseteq \mathcal O_i \times \mathcal O_j$ a relational context, $\rho$ a scaling operator, $o \in \mathcal O_i$ an object, $OC_j$ the set of object-concepts of $\mathcal K_j = (\mathcal O_j, \mathcal A_j, \mathcal I_j)$}
\KwOut{Extends the context $\mathcal K_i$ and adds the crosses}
\If{$\rho==\exists$}{
	\ForEach{$(X,Y)\in OC_j$ such that $\exists obj\in r(o)$, $obj\in X$ }{
		$\mathcal A_i\gets\mathcal A_i\cup \exists r.(X,Y)$\;
		$\mathcal I_i\gets\mathcal I_i\cup (o, \exists r.(X,Y))$\;
	}
}

\If{$\rho==\exists\forall$}{
%	$X\gets\mathcal A_j$\;
  $X\gets ${\sc In}$(\mathcal{K}_j,r(o))$

%	\ForEach{$obj\in r(o)$}{
%		$X\gets ${\sc Intersect}$(X, ${\sc ~In}$(\mathcal K_j,OC_j(obj)))$\;
%    }
    $\mathcal A_i\gets\mathcal A_i\cup \exists\forall r.(${\sc Ex}$(\mathcal K_i, X), X)$\;
	$\mathcal I_i\gets\mathcal I_i\cup (o, \exists\forall r.(${\sc Ex}$(\mathcal K_i,X), X))$\;
}
\caption{{\sc GrowContext}$(\mathcal K_i,r,q,o,OC_j)$}
\label{algo:Extends}
\end{algorithm}
%3 pages
\section{Example}
\label{sec:examples}
In this section, we illustrate the defined algorithms.
We consider the RCF $(\textbf{K}_s,\textbf{R}_s)$ with $\textbf{K}_s = \{DM\_tools, DBMS\}$ and $\textbf{R}_s = \{support\}$ as presented in Section~\ref{sec:rca}.
We decide to
%associate the existential scaling operator to \textit{support}, thus defining
 apply the strategy $\{(support,\exists)\}$.

Let us imagine that a user wants to select a data modelling tool that runs on Windows (\textit{OS:Windows}) and that handles logical and conceptual data models (\textit{DM:Logical} and \textit{DM:Conceptual}).
Traditional FCA may compute the formal concept associated with these 3 attributes (i.e., \textit{C\_DM\_tools\_5}, left-hand side of Fig.~\ref{fig:conceptlattices}), and inform the user that 1) the corresponding tools are \texttt{Erwin DM}, \texttt{Magic Draw} and \texttt{ER/Studio}, and that 2) all these tools also handle \textit{DM:Physical}.

Let us apply our algorithms on this concept to 1) retrieve the supported DBMS  (relational cover) and 2) find the closest alternatives to the query (lower and upper covers): {\sc RCA}($\textbf{K}_s, \{(support,\exists)\}, C\_DM\_tools\_5, DM\_tools)$.
%With RCA, the user may also be informed of what DBMS are supported by the queried data modelling tools.
%We applied the {\sc RCA} algorithm, using the existential scaling operator in our strategy to retrieve this information: {\sc RCA}($\textbf{K}_s, \{(support,\exists)\}, C\_DM\_tools\_5, DM\_tools)$.

Lines 1 to 4 extend the context of \textit{DM\_tools} with the relational attributes representing the object-concepts of \textit{DBMS} (\textit{support}'s target context).
In our case, we have only one relation (\textit{support},$~\exists$) visited at Line 1.
%Line 1, one loop is necessary because we have only one strategy (\textit{support},$\exists$).
In Line 2, $OC_j$ takes the object-concepts of \textit{DBMS} , i.e., concepts 1, 2, 3 and 4 from the right-hand side of Fig.~\ref{fig:conceptlattices}.
Then, the loop on Lines 3 and 4 considers the 5 objects of \textit{DM\_tools}, on which {\sc GrowContext} is called.
%In the later, only the case "$\exists$" is visited, following our strategy.
Each object $o_i$ of \textit{DM\_tools} is associated to the relational attributes representing the concepts of $OC_j$ having in their extents at least one object linked with $o_j$. %introducing the \textit{DBMS} they are linked with.

As $support(\texttt{Astah}) = \{MySQL,Oracle\}$,  $\exists support(C\_DBMS\_3)$ (\textit{MySQL} object-concept) and $\exists support(C\_DBMS\_2)$ (\textit{Oracle} object-concept) are added to \textit{DM\_tools} and associated to \texttt{Astah}.
%In the same way, \texttt{Erwin DM} is associated to these two relational concepts, and add $\exists support(C\_DBMS\_1)$ (\texttt{Teradata} object-concept) to the extended context.
At the end of Line 4, we obtain the extended context presented in Table~\ref{table:rcf-extendedformalcontext}.
%\jc{Comme Erwin DM est en relation avec MySQL, Oracle et Terradata, il est associe aux concepts 1, 2 et 3 de DBMS. Or, il devrait aussi etre associe au concept 4, qui introduit Postgre (objet non lié à Erwin DM), mais qui a aussi dans son extent Terradata.}
\begin{table}[ht]
  \scriptsize
  \caption{Formal context \textit{DM\_tools} extended according to the relation \textit{support}}
  \label{table:rcf-extendedformalcontext}
  \centering
  \begin{tabular}{|l|c|c|c|c|c|c|c||c|c|c|c|}
    \hline
    \textit{DM\_tools$^+$}&\begin{sideways}OS:Windows\end{sideways}&\begin{sideways}OS:Mac OS\end{sideways}&\begin{sideways}OS:Linux\end{sideways}&\begin{sideways}DM:Conceptual~~\end{sideways}&\begin{sideways}DM:Physical\end{sideways}&\begin{sideways}DM:Logical\end{sideways}&\begin{sideways}DM:ETL\end{sideways}
    &\begin{sideways}${\exists}$ sup.(C\_DBMS\_1)\end{sideways}
    &\begin{sideways}${\exists}$ sup.(C\_DBMS\_2)\end{sideways}
    &\begin{sideways}${\exists}$ sup.(C\_DBMS\_3)\end{sideways}
    &\begin{sideways}${\exists}$ sup.(C\_DBMS\_4)\end{sideways}\\
    \hline
    Astah&x&x&x&x&&&&&x&x&\\
    \hline
    Erwin DM&x&&&x&x&x&&x&x&x&x\\
    \hline
    ER/Studio&x&&&x&x&x&x&x&x&x&x\\
    \hline
    Magic Draw&x&x&x&x&x&x&&&x&x&x\\
    \hline
    MySQL Workbench&x&x&x&&x&&&&&x&\\
    \hline
    \end{tabular}
  \end{table}

Line 5 updates the intent of the input concept to take into account the relational attributes: $\{$\textit{OS:Windows, DM:Conceptual, DM:Physical, DM:Logical, $\exists sup.(C\_DBMS\_2)$, $\exists sup.(C\_DBMS\_3)$, $\exists sup.(C\_DBMS\_4)$}\}).
The concepts of \textit{DBMS} corresponding to the relational attributes of $C$ ($C\_DBMS\_2$ to $4$) form the relational cover of the input concept (lines 6 to 8).
%\jc{Puisque Erwin n'était pas associé a l'attribut relationel representant le concept 4 de DBMS, C\_DBMS\_4 n'est pas dans le nouvel extent de C, alors qu'il devrait (voir Fig 2).}

Then, (Lines 9 to 11) we compute the minimal generators of the extent of $C$, which are $\{$\texttt{Erwin DM}, \texttt{Magic Draw}$\}$ and $\{$\texttt{ER/Studio, Magic Draw}$\}$.
Their minimal transversals are $\{\texttt{Magic~Draw}\}$ and $\{\texttt{Erwin~DM, ER/Studio}\}$. The two concepts having $\{\texttt{Magic~Draw}\}$ and $\{\texttt{Erwin~DM,ER/Studio}\}$ for extent represent the lower cover (respectively $C\_DM\_tools\_1$ and $C\_DM\_tools\_2$ in Fig.~\ref{fig:extendedconceptlattice}).

Finally, in Lines 12 to 14, we consider the objects of \textit{DM\_tools} that are not in $C$'s extent, i.e.,  \texttt{MySQL Workbench} and \texttt{Astah}.
For each one of them, we compute the concept corresponding to their union with $C$'s extent, and we obtain the two concepts \textit{C\_DM\_tools\_7} and \textit{8} of Fig.~\ref{fig:extendedconceptlattice}.
They represent the upper cover of $C$.

\section{Related Work}
\label{sec:relatedwork}

%% Exploratory search and FCA
Lattice structures are among the first structures used as a support for exploratory search \cite{Godin1986Browsable},
and this task has later attracted a lot of attention in Formal Concept Analysis theory \cite{Codocedo2015Survey}.
Many works focus on \textit{conceptual neighbourhood} to present both information related to a query and its closest variants \cite{Godin1989Design,Ducrou2007,Alam2016}.
In this paper, we consider RCA to retrieve the conceptual neighbourhood in interconnected lattices, structuring both intrinsic and relational attributes.

%% Reducing the complexity of FCA-based exploratory search
The exponential growth of concept lattices is well-known \cite{Ganter1999Formal}.
As a consequence, the main limitation of FCA-based exploratory search lies in the complexity and computation of the structures \cite{Carpineto2004}.
Many solutions have been proposed to reduce the complexity of conceptual navigation.
Some authors propose to prune the concept lattice to restrict the explorable dataspace, by computing iceberg concept lattices \cite{Stumme2002}, or by applying constraints to bound the final structure \cite{Carpineto2004}.
To ease the navigation, the authors of \cite{Melo2013Browsing} seek to extract more simplified browsable structures; they first extract a tree from the concept lattice, and then reduce the obtained tree using clustering and fault-tolerance methods.
The tool \texttt{SearchSleuth} \cite{Ducrou2007} enables FCA-based exploratory search for web queries, a field where the domain cannot be entirely processed using FCA and concept lattices.
To tackle this issue, they generate a new formal context specific to a query at each navigation step. %, from which they compute the corresponding conceptual neighbourhood.
In a previous work \cite{Bazin2017Generation}, we proposed to compute the conceptual neighbourhood of a query in a sub-order of the concept lattice restricted to the attribute- and object-concepts (attribute-object-concept poset), a condensed alternative to concept lattices.
At each step, only the conceptual neighbourhood is computed.
In the present work, we also generate the conceptual neighbourhood on-the-fly, but this time in interconnected concept lattices.

%% RCA to navigate in connected documents

Mimouni et al. \cite{Mimouni2015Legal} use RCA to structure, query and browse a collection of legal documents.
First, they build interconnected lattices representing different types of legal documents referring to each other.
Then, their approach allows for the retrieval of the concept corresponding to a user query, and to explore variations of this query by navigation in the neighbour concepts.
In their approach, they compute all the lattices during the first step. %, but assert that the main limitation is the cost of this computation.
%They outline some solutions as using iceberg lattices to reduce the complexity, or to prune the initial dataset before the computation.

Ferr\'e and Hermann \cite{ferre2012reconciling} propose \textit{Query-based Faceted Search} and an implementation in the tool \texttt{SEWELIS}, that allows to browse relational datasets in the form of RDF files.
Also, Ferr\'e et al. \cite{ferre2005arbitrary} propose RLCA, a relational extension of \textit{Logical Formal Analysis}, an adaptation of FCA to describe objects by formulas of ad-hoc logics instead of binary attributes.
While RCA computes connected yet separate concept lattices, one per sort of objects, RLCA gathers the objects, their descriptions and their relations to other objects in one structure.
%1 pages (refs comptees ci-dessous)
\section{Conclusion}
\label{sec:conclusion}

In this paper,
we proposed algorithms to compute the conceptual neighbourhood of a query in connected concept lattices generated with RCA.
First, we redefined the traditional FCA derivation operators to take into account relational attributes.
Then, we presented a way to compute the relational, upper and lower covers of a given concept in extended lattices, without computing all the structures.
Two RCA scaling operators, i.e., existential and universal strict, may be used.
We illustrated how the algorithms work on a running example from the domain of software product line engineering.

In the future, we plan to study the properties of the algorithm and to implement it to perform exploratory search in relational datasets.
A scalability study on real datasets from the projects Fresqueau and Knomana and from available product descriptions \cite{bennasr:hal-01427218} is then envisioned. To this end, we will generate random queries and exploration paths. We also are collecting concrete questions from the Knomana project partners for having real exploration tasks in their domain and qualitatively evaluate the benefits of the approach.
%2 pages avec la biblio

\bibliographystyle{splncs03}
\bibliography{biblio}

\end{document}